\documentclass[twocolumn,
pra, 
 showpacs,floats,preprintnumbers,amsmath,amssymb]{revtex4}

\usepackage{epsfig}
\usepackage{amsmath}
\renewcommand{\(}{\left(}
\renewcommand{\)}{\right )}
\renewcommand{\[}{\left [}
\renewcommand{\]}{\right ]}

\def\bea{\arraycolsep .1em \begin{eqnarray}}
\def\eea{\end{eqnarray}}

\let\om=\omega
\let\be=\beta
\let\no=\nonumber

\def\eq#1{Eq.(\ref{#1})}
\def\refr#1{\cite{#1}}

\def\s0#1#2{\mbox{\small{$ \frac{#1}{#2} $}}}
\def\0#1#2{\frac{#1}{#2}}

\def\pra#1#2#3{Phys. Rev.  {\bf A #1}, #2 (#3)}

\def\prl#1#2#3{Phys. Rev. Lett. {\bf #1}, #2 (#3)}

\begin{document}

\title{Comparative study of the finite-temperature thermodynamics of a
unitary Fermi gas}
\author{Fang Qin\footnote{Email:
qinfang@phy.ccnu.edu.cn} and Ji-sheng Chen\footnote{Email:
chenjs@iopp.ccnu.edu.cn }} \affiliation{Physics Department $\&$
Institute of Particle Physics, Central China Normal University,
Wuhan 430079, People's Republic of China}
\begin{abstract}
We study the finite-temperature thermodynamics of a unitary Fermi
gas. The chemical potential, energy density and entropy are given
analytically with the quasi-linear approximation. The ground state
energy agrees with previous theoretical and experimental results.
Recently, the generalized exclusion statistics is applied to the
discussion of the finite-temperature unitary Fermi gas
thermodynamics. A concrete comparison between the two different
approaches is performed. Emphasis is made on the behavior of the
entropy per particle. In physics, the slope of entropy gives the
information for the effective fermion mass $m^*/m$ in the low
temperature strong degenerate region. Compared with $m^*/m \approx
0.70<1$ given in terms of the generalized exclusion statistics, our
quasi-linear approximation determines $m^*/m\approx 1.11>1$.

{\bf Keywords}:Unitary Fermi gas thermodynamics; Quasi-linear
approximation method; Generalized exclusion statistics
\preprint{0904.0053}
\end{abstract}
\pacs{12.40.Ee; 05.30.Pr; 05.70.-a}

\maketitle

\section{Introduction}

In recent years, the strongly interacting fermion physics becomes
the focus of theoretical and experimental
attention\cite{Giorgini2008}. This is much attributed to the rapid
progress of the atomic Fermi gas experiments.

By tuning the external magnetic field, one can control the $S$-wave
scattering length $a$ or interaction strength between two atomic
fermions. The crossover from Bardeen-Cooper-Schrieffer (BCS) to
Bose-Einstein condensation (BEC) can be realized by the so-called
Feshbach resonance\cite{Science2004}. At the resonance point, the
scattering length can be singular with the existence of a
zero-energy bound sate. Although the scattering length is singular,
the scattering cross-section is saturated as $\sigma \sim 4\pi/k^2$
(with $k$ being the relative momentum between two atomic fermions)
due to the unitary property limit. The divergent scattering fermion
thermodynamics is referred to as the unitary Fermi gas
thermodynamics in the literature\cite{Ho2004}. Dealing with the
strongly interacting matter is related with a variety of realistic
many-body topics.

Usually, the thermodynamics of dilute fermion system is determined
by the two-body scattering length $a$, particle number density $n$
and temperature $T$. In the unitary limit with $a=\pm \infty$, the
dynamical scattering limit should drop out in the thermodynamic
quantities. At unitarity, the dynamical detail should not affect the
thermodynamics; i.e., the unitary fermion system can manifest the
universal properties\cite{Ho2004}.

Due to lack of any small expansion parameter, the unitary Fermi gas
provides an intractable problem in statistical physics. The
fundamental issue is on the zero-temperature ground state energy.
Based on the dimensional analysis, the ground state energy should be
proportional to that of the ideal Fermi gas with a universal
constant $\xi =1+\be$, which excites many theoretical and
experimental efforts. The world average value of $\xi$ is
$0.42-0.46$\cite{constant1,constant2,constant3,constant4,constant5}.
Recently, we have attempted a quasi-linear approximation method to
explore the strongly interacting limit fermion
thermodynamics\cite{Chenjs}. The obtained ground state energy or the
universal constant $\xi=\049$ is reasonably consistent with some
theoretical or experimental investigations.

Generally, the finite-temperature thermodynamics is as intriguing as
the zero-temperature ground state energy. There have been several
Monte Carlo finite-temperature calculations of a unitary Fermi
gas\cite{MC1,MC2}. In the strongly correlation unitary fermions, the
nonlinear quantum fluctuations/correlations compete with dynamical
high order effects. In the weak degenerate Boltzmann regime, the
nonlinear correlations make the second order virial coefficient
$a_2$ vanish. To a great extent, the vanishing leading order quantum
correction reflects the \textit{intermediate} crossover
characteristics of a unitary Fermi gas\cite{Chenjs}.

Can the intermediate characteristics be described in another way? In
\cite{Haldane1991,Wu1994}, the generalized exclusion statistics was
developed to describe the anyon behavior in the low-dimensional
strongly correlation quantum system. Physically, the behavior of a
unitary Fermi gas is between Bose gas and Fermi gas\cite{MC1}.
Similarly, the behavior of anyons is also between bosons and
fermions. Can one use the anyons statistics to describe the
intermediate unitary Fermi gas? Recently, the generalized exclusion
statistics has been generalized to describe the unitary Fermi gas
thermodynamics\cite{Bhaduri1,Bhaduri2}. As a hypothesis, the
priority is that the thermodynamics at finite-temperature can be
investigated quantitatively.

From the general viewpoint of statistical mechanics, calculating
entropy is not a simple task. Either in classical or quantum theory,
the entropy describes how the microscopic states are counted
properly. From the quantum degenerate viewpoint, the low-temperature
behavior of the entropy is a characteristic quantity. For example,
according to the Landau theory for the strong correlation
Fermi-Liquid, the slope of entropy per particle versus temperature
is related with the effective fermion mass $m^*/m$. In physics, the
dynamical parameter $m^*/m$ is very important for the phase
separation discussion for the asymmetric fermion system with unequal
populations\cite{Lobo2006,Pilati2008,Combescot2007}. Like the
universal constant $\xi =1+\be$, the effective fermion mass $m^*/m$
is an another universal constant for the BCS-BEC crossover
thermodynamics. Obviously, the physics beyond the mean-field theory
should be reasonably well understood.

Unlike the ground state energy or the universal constant $\xi$ with
the world average value $\xi\approx 0.44$, the effective fermion
mass is an unknown parameter up to now. For example, the effective
fermion mass is estimated to be $m^*/m\approx 1.04$ with a quantum
Monte Carlo calculation\cite{Lobo2006}. A quantitative study of the
phase diagram at zero temperature along the BCS-BEC crossover using
fixed-node diffusion Monte Carlo simulations shows $m^*/m\approx
1.09$\cite{Pilati2008}. A many-body variational wave function with a
T-matrix approximation leads to a larger value $m^*/m \approx
1.17$\cite{Combescot2007}. What is the exact value of $m^*/m$?

In a quantitative way, we make a comparative study for the
finite-temperature thermodynamic properties of the unitary fermion
gas with the two formulations. The behavior of entropy per particle
based on the quasi-linear approximation and the generalized
exclusion statistics is discussed in detail. Indirectly, the
effective fermion mass is determined from the entropy. The results
are further compared with the Monte Carlo calculations.

The paper is organized in the following way. In Sec.\ref{section2},
the relevant thermodynamic expressions are given by the quasi-linear
approximation. Correspondingly, the thermodynamics given by the
generalized exclusion statistics is presented in Sec.\ref{section3}.
The numerical calculations and concrete comparisons between the two
methods are given in Sec.\ref{section4}. In this section, the
entropy per particle and corresponding effective fermion mass
$m^*/m$ are discussed. In Sec.\ref{section5}, we present the
conclusion remarks.

\section{Thermodynamical quantities given by statistical dynamics with quasi-linear approximation}\label{section2}

Strongly correlated matter under extreme conditions often requires
the use of effective field theories in the description for the
thermodynamic properties, independently of the energy scale under
consideration. In the strongly interacting system, the central task
is how to deal with the non-perturbative fluctuation and correlation
effects. In Ref.\cite{Chenjs}, a quasi-linear approximation is taken
to account for the non-local correlation effects on the unitary
Fermi gas thermodynamics.

With the quasi-linear approximation method, the obtained grand
thermodynamic potential $\Omega (T,\mu) $ or pressure $P=-\Omega/V$
can be described by the two coupled-parametric equations through the
intermediate variable-effective chemical potential $\mu^*$
 \bea\label{1}
P=\frac{2T}{\lambda^{3}}f_{{5}/{2}}(z^{'})+\frac{\pi
a_{eff}}{m}n^{2}+n\mu_{r},\eea \bea\label{2} \mu=\mu^{*}+\frac{2\pi
a_{eff}}{m}n+\mu_{r}. \eea

In the above equations, $\lambda=\sqrt{{2\pi}/({mT})}$ is the
thermal de Broglie wavelength and $m$ is the bare fermion mass(with
natural units $k_{B}=\hbar=1$ throughout the paper).

The effective chemical potential $\mu^{*}$ is introduced by the
single-particle self-consistent equation. $\mu^*$ makes the
thermodynamic expressions appear as the standard Fermi integral
formalism\bea
f_{\upsilon}(z^{'})=\frac{1}{\Gamma(\upsilon)}\int_{0}^{\infty}{\frac{x^{\upsilon-1}dx}{z^{'-1}e^{x}+1}},
\eea where $\Gamma(\upsilon)$ is the gamma function, and
$z^{'}=e^{{\mu^{*}}/{T}}$ is the effective fugacity. For example,
the quasi-particle Fermi-Dirac distribution function gives the
particle number density according to \bea
n=\frac{2}{\lambda^{3}}f_{{3}/{2}}(z^{'}).\eea In the coupled
equations \eq{1} and \eq{2}, the shorthand notations are defined
as\bea a_{eff}=-\frac{m}{2\pi m_{D}^{2}},~~~~~~
m_{D}^{2}=\(\frac{\partial n}{\partial \mu^{*}}\)_{T}.\eea

The shift term $\propto\mu_r$ characterizes the high order
    nonlinear contributions, which strictly ensures the energy-momentum conservation law.
In the nonlinear approximation, this significant high order
correction term can be fixed in a thermodynamic way. It is worthy
noting that the term $\propto\mu_r$ can be exactly canceled by each
other in the Helmholtz free energy density \bea
\frac{F}{V}=f=-P+n\mu, \eea where $V$ is the system volume. However,
the high order correlation term $\propto\mu_r$ can be obtained in
terms of the thermodynamic relations\cite{Chenjs} \bea\label{temp1}
P=-\(\frac{\partial F}{\partial
V}\)_{T,N}=-\(\frac{\partial\frac{F}{N}}{\partial\frac{V}{N}}\)_{T}=n^{2}\(\frac{\partial\frac{f}{n}}{\partial
n}\)_{T}, \eea and \bea\label{temp2} \mu=\(\frac{\partial
F}{\partial
N}\)_{T,V}=\(\frac{\partial\frac{F}{V}}{\partial\frac{N}{V}}\)_{T}=\(\frac{\partial
f}{\partial n}\)_{T}. \eea Comparing those obtained from \eq{temp1}
and \eq{temp2} with \eq{1} and \eq{2}, the explicit expression of
$\mu_r$ is\bea \mu_{r}=\frac{1}{2}\(\frac{\partial
m_{D}^{2}}{\partial n}\)_{T}\(\frac{2\pi a_{eff}}{m}\)^{2}n^{2},
\eea

The integrated expressions of the pressure and chemical potential
for the unitary Fermi gas are \bea\label{pressure}
P&&=\frac{2T}{\lambda^{3}}\(f_{{5}/{2}}(z^{'})-
\frac{f_{{3}/{2}}^{2}(z^{'})}{2f_{{1}/{2}}(z^{'})}+
\frac{f_{{3}/{2}}^{3}(z^{'})f_{-{1}/{2}}(z^{'})}{2f_{{1}/{2}}^{3}(z^{'})}\),\no\\
\\
\label{3}
\mu&&=\mu^{*}-T\frac{f_{{3}/{2}}(z^{'})}{f_{{1}/{2}}(z^{'})}
+\frac{T}{2}\frac{f_{{3}/{2}}^{2}(z^{'})f_{-{1}/{2}}(z^{'})}{f_{{1}/{2}}^{3}(z^{'})}.
\eea

In the quasi-linear approximation, the auxiliary implicit variable
$\mu^*$ is introduced to characterize the non-linear
fluctuation/correlation effects.  As indicated by \eq{pressure} and
\eq{3}, the $\mu^*$ or $z'$ makes the realistic grand thermodynamic
potential $\Omega (T,\mu)$ appear as the set of highly non-linear
parametric equations, which can be represented by the standard Fermi
integral.
 By eliminating the auxiliary variable $\mu^*$, the equation of
state will uniquely be determined.

From the underlying grand thermodynamic potential-partition
function, one can derive the analytical expressions for the entropy
density $s=S/V$ and internal energy density $\epsilon=E/V$. The
following partial derivative formulae will be used \bea
&&\(\frac{\partial\mu^{*}}{\partial T}\)_{\mu}\(\frac{\partial
T}{\partial
\mu}\)_{\mu^{*}}\(\frac{\partial\mu}{\partial\mu^{*}}\)_{T}=-1, \no\\
&&\(\frac{\partial m_{D}^{2}}{\partial T}\)_{n}=\(\frac{\partial
m_{D}^{2}}{\partial T}\)_{\mu^{*}}+\(\frac{\partial
m_{D}^{2}}{\partial\mu^{*}}\)_{T}\(\frac{\partial \mu^{*}}{\partial
T}\)_{n}.\eea

The entropy is derived according to \bea\label{4}
&&\frac{s}{n}=\frac{1}{n}\(\frac{\partial P}{\partial T}\)_{\mu}\no\\
&&=\frac{5}{2}\frac{f_{{5}/{2}}(z^{'})}{f_{{3}/{2}}(z^{'})}-\ln{z^{'}}+
\frac{3f_{-{1}/{2}}(z^{'})f_{{3}/{2}}^{2}(z^{'})}{4f_{{1}/{2}}^{3}(z^{'})}
-\frac{f_{{3}/{2}}(z^{'})}{4f_{{1}/{2}}(z^{'})}.\no\\ \eea
Correspondingly, the explicit energy density expression is
calibrated to be \bea\label{5}
\epsilon=\frac{3T}{\lambda^{3}}\(f_{{5}/{2}}(z^{'})-
\frac{f_{{3}/{2}}^{2}(z^{'})}{2f_{{1}/{2}}(z^{'})}+
\frac{f_{{3}/{2}}^{3}(z^{'})f_{-{1}/{2}}(z^{'})}{2f_{{1}/{2}}^{3}(z^{'})}\).\no\\
\eea

Essentially, the entropy density includes the high order nonlinear
contribution. What we want to emphasize is that the third law of
thermodynamics is exactly ensured as expected.  The analytical
analysis indicates that the energy density at zero-temperature gives
the dimensionless universal coefficient according to
$\xi={\mu}/{E_{F}}=\049$ or ${E}/{(\035 NE_{F})}=\xi $, where the
Fermi energy is $E_{F}={(3\pi^{2}n)^{{2}/{3}}}/{(2m)}$ and $T_{F}$
is the Fermi characteristic temperature in the unit Boltzmann
constant. The universal coefficient $\xi=\049 $ has attracted much
attention in the literature and is reasonably consistent with some
Monte Carlo calculations\cite{constant1,MC1}.

\section{Thermodynamics given by the generalized exclusion
statistics}\label{section3}

\subsection{Generalized exclusion
statistics}

The generalized exclusion statistics is proposed in
\cite{Haldane1991} and \cite{Wu1994}. If the dimensional of the
Hilbert space is $d$ and the particle number is $N$, then $d$ and
$N$ are connected by $\bigtriangleup d=-g\bigtriangleup N$, where
the shift of the single-particle states number is $\bigtriangleup
d$. The shift of the particle number for identical particle system
is $\bigtriangleup N$ and $g$ is a statistical parameter, which
denotes the ability of one particle to exclude other particles in
occupying single-particle state. When $g=0$ the intermediate
statistics returns to the Bose-Einstein statistics and $g=1$ to the
Fermi-Dirac statistics.

For anyons, the number of quantum states $W$ of $N$ identical
particles occupying a group of $G$ states are determined by the
interpolated statistical weights of the Bose-Einstein and
Fermi-Dirac statistics. A simple formula with the generalized
exclusion statistics is used to describe the microscopic quantum
states\cite{Wu1994} \bea\label{W1}
W=\frac{[G+(N-1)(1-g)]!}{N![G-gN-(1-g)]!}. \eea

One can divide the one-particle states into a large number of cells
with $G\gg1$ states in each cell, and calculate the number with
$N_i$ particles in the $i$-th cell. The total energy and the total
number of particles are fixed and given as \bea
E=\sum_{i}N_{i}\epsilon_{i}, N=\sum_{i}N_{i}, \eea with
$\epsilon_{i}$ defined as the energy of particle of species $i$. By
generalizing \eq{W1}, we have \bea
W=\prod_{i}\frac{[G_{i}+(N_{i}-1)(1-g)]!}{N_{i}![G_{i}-gN_{i}-(1-g)]!}.
\eea

We consider a grand canonical ensemble at temperature $T$. For very
large $G_i\gg1$ and $N_i\gg1$, using the Stirling formula
$\ln{N!}=N(\ln{N}-1)$, and introducing the average occupation number
defined by $\bar{N_i}\equiv N_i/G_i$, one has \bea\label{W2}
\ln{W}&&=\sum_{i}\ln{\[\frac{[G_{i}+(N_{i}-1)(1-g)]!}{N_{i}![G_{i}-gN_{i}-(1-g)]!}\]}\no\\
&&\simeq\sum_{i}\[G_i\(1+(1-g)\bar{N_i}\)\ln{G_i\(1+(1-g)\bar{N_i}\)}
\right. \no\\
&& \left.
~~-G_i(1-g\bar{N_i})\ln{G_i(1-g\bar{N_i})}-G_i\bar{N_i}\ln{G_i\bar{N_i}}\].\no\\
\eea

Through the Lagrange multiplier method, the most probable
distribution of $\bar{N_i}$ is determined by \bea
\frac{\partial}{\partial\bar{N_i}}[\ln{W}-\sum_{i}G_i\bar{N_i}(\epsilon_i-\mu)/T]=0,
\eea with chemical potential $\mu$. It follows that \bea
\bar{N_i}e^{(\epsilon_i-\mu)/T}=[1+(1-g)\bar{N_i}]^{(1-g)}(1-g\bar{N_i})^g.
\eea Setting $\omega_i=1/\bar{N_i}-g$, we have the anyon statistical
distribution \bea\label{N1} \bar{N_i}=\frac{1}{\omega_i+g}, \eea
where $\omega$ obeys the relation \bea\label{10}
\omega^{g}(1+\omega)^{1-g}=e^{(\epsilon-\mu)/T}. \eea

One can define $\omega_{0}$ of $\omega$ at $\epsilon=0$ with \eq{10}
\bea\label{11} \mu=-T\ln{[\omega_{0}^{g}(1+\omega_{0})^{1-g}]}. \eea
The relation between $\mu$ and $T$ has been established indirectly
through $\omega_0$ and $g$. From \eq{10}, the $\om$ and $\omega_0$
are related with each other through single-particle energy
$\epsilon$\bea\label{12}
\epsilon=T\ln{\[\(\frac{\omega}{\omega_{0}}\)^{g}\(\frac{1+\omega}{1+\omega_{0}}\)^{1-g}\]},
\eea which gives \bea\label{13}
d\epsilon=\frac{T(g+\omega)}{\omega(1+\omega)}d\omega. \eea

For $T=0$, the average occupation number can be explicitly indicated
as \bea\label{14} \bar{N}=\left\{
            \begin{array}{ll}
              0, & \hbox{     if $\varepsilon>\mu$;} \\
              \frac{1}{g}, & \hbox{     if $\varepsilon<\mu$,}
            \end{array}
          \right.
  \eea
which is quite similar to the Fermi-Dirac statistics.

\subsection{Particle number and energy densities}

In the anyon statistics, the density of states is also given by \bea
D(\epsilon)=\alpha{(2m)^{3/2}}V\epsilon^{1/2}/({4\pi^{2}}), \eea
where $\alpha$ is the degree of the spin degeneracy and $m$ is the
bare fermion mass.

At $T=0$, the particle number is explicitly given by \bea\label{a}
N=\frac{1}{g}\int_{0}^{\widetilde{E}_{F}}{D(\epsilon)d\epsilon}=\frac{\alpha(2m)^{{3}/{2}}}{6\pi^{2}}VE_{F}^{{3}/{2}},
\eea where $\widetilde{E}_{F}$ is related with the Fermi energy
$E_{F}$ through $\widetilde{E}_{F}=g^{{2}/{3}}E_{F}$. With the
$\widetilde{E}_{F}$ symbol, the system energy can be represented
as\bea E=\frac{1}{g}\int_{0}^{\widetilde{E}_{F}}{\epsilon
D(\epsilon)d\epsilon}=\frac{3}{5}g^{{2}/{3}}NE_{F}. \eea

As we will see, once $g$ is fixed, one can discuss the general
finite-temperature thermodynamic properties. Therefore, the
essential task in the generalized exclusion statistics is fixing the
statistical factor $g$. This can be determined by the
zero-temperature ground state energy or the universal constant $\xi$
according to $\xi=g^{{2}/{3}}$. Various theoretical or experimental
attempts have been made in the literature for determining the ground
state energy. With the universal coefficient
$\xi=\049$\cite{Chenjs}, the expected statistical factor can be
identified to be $g=\frac{8}{27}$.

For the general finite-temperature scenario, the particle number and
energy can be rewritten as \bea\label{18}
N=\int_{0}^{\infty}{\frac{D(\epsilon)d\epsilon}{\omega+g}},\\
\label{19} E=\int_{0}^{\infty}{\frac{\epsilon
D(\epsilon)d\epsilon}{\omega+g}}. \eea

By replacing \eq{12}-\eq{13} and \eq{a} into \eq{18} and \eq{19},
one can have \bea\label{20}
\frac{3}{2}&&\(\frac{T}{T_{F}}\)^{3/2}a(\omega_{0})=1,\\
\label{21}
\frac{E}{NE_{F}}&&=\frac{3}{2}\(\frac{T}{T_{F}}\)^{5/2}b(\omega_{0}),\\
a(\omega_{0})&&=\int_{\omega_{0}}^{\infty}{\frac{d\omega}{\omega(1+\omega)}
\[\ln{\(\frac{\omega}{\omega_{0}}\)^{g}\(\frac{1+\omega}{1+\omega_{0}}\)^{1-g}}\]^{1/2}},\no\\
b(\omega_{0})&&=\int_{\omega_{0}}^{\infty}{\frac{d\omega}{\omega(1+\omega)}
\[\ln{\(\frac{\omega}{\omega_{0}}\)^{g}\(\frac{1+\omega}{1+\omega_{0}}\)^{1-g}}\]^{3/2}}.\no\eea

\eq{20} determines $\omega_{0}$ for a given temperature $T$.
${E}/{(NE_{F})}$ can be obtained by a given $\omega_{0}$ through
\eq{21}.

For giving the explicit entropy density expression with the
generalized exclusion statistics in the next subsection, let us make
further discussion for the energy density. By eliminating $N$ with
\eq{a} and \eq{21}, the energy can be alternatively expressed as
\bea
E=\frac{\alpha(2m)^{{3}/{2}}}{4\pi^{2}}VT^{{5}/{2}}b(\omega_{0}).
\eea The partial derivative of the internal energy $E$ to $T$ for
fixed $\mu$ is given by \bea\label{e} \(\frac{\partial E}{\partial
T}\)_{\mu}=\frac{\alpha
V(2m)^{{3}/{2}}}{4\pi^{2}}T^{{3}/{2}}\[\frac{5}{2}b(\omega_{0})+
T\(\frac{\partial b(\omega_{0})}{\partial T}\)_{\mu}\].\no\\ \eea

Furthermore, the variable $\omega_{0}$ of the integral function
$b(\omega_{0})$ can be converted into $\mu$ and $T$ through \eq{11}
\bea\label{above}
b(\omega_{0},\mu,T)=\int_{\omega_{0}}^{\infty}{\frac{d\omega}{\omega(1+\omega)}
\[\ln(\omega^{g}(1+\omega)^{1-g})+\frac{\mu}{T}\]^{{3}/{2}}}.\no\\
\eea Therefore, one can have \bea\label{f} \(\frac{\partial
b}{\partial
T}\)_{\mu}=\frac{3}{2T}\ln[\omega_{0}^{g}(1+\omega_{0})^{1-g}]a(\omega_{0}).
\eea

\subsection{Entropy per particle}

Due to the scaling properties, the thermodynamics of a unitary Fermi
gas also satisfies the ideal gas virial theorem
\cite{Ho2004,Chenjs,Thomas2005} \bea\label{b}
P=\frac{2}{3}\frac{E}{V}. \eea

According to the thermodynamic relation for the entropy $S$ and
pressure $P$, one can have\bea\label{d}
S=\frac{2}{3}\(\frac{\partial E}{\partial T}\)_{\mu}. \eea

By substituting \eq{e} and \eq{f} into \eq{d}, the explicit
expression for the entropy per particle is derived to
be\bea\label{g}
\frac{S}{N}=\frac{5}{2}\(\frac{T}{T_{F}}\)^{{3}/{2}}b(\omega_{0})+\ln[\omega_{0}^{g}(1+\omega_{0})^{1-g}],
\eea where $\omega_{0}$ is given by \eq{20} for a given $T$.

\section{Numerical results and comparisons}\label{section4}

Based on the above analytical expressions, we will give the
numerical results.

\subsection{Internal energy and chemical potential}

\begin{figure}[ht]
        \centering
        \psfig{file=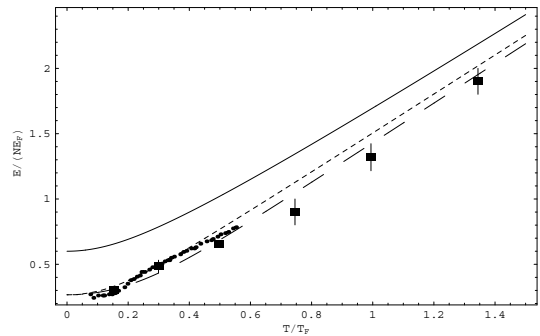,width=7cm,angle=-0}
        \caption{The internal energy per particle versus the rescaled temperature.
        The solid curve denotes that for the ideal
Fermi gas, and the short-dashed one is that given by the
quasi-linear approximation. The long-dashed curve represents the
result in terms of the generalized exclusion statistics model. The
dots and solid squares are the Monte Carlo calculations \cite{MC1}
and \cite{MC2}, respectively.
        \small
}\label{fig1}
\end{figure}

From \eq{20} and \eq{21}, the energy per particle versus the
rescaled temperature can be solved. As indicated by Fig.\ref{fig1},
the internal energies for the unitary Fermi gas based on the
quasi-linear approximation and the generalized exclusion statistics
have similar analytical properties; i.e., the internal energy
increases with the increase of temperature. The two approaches both
show that the energy density of a unitary Fermi gas is lower than
that of the ideal Fermi gas. However, the shift of the internal
energy given by the quasi-linear approximation is more quicker than
that determined by the generalized exclusion statistics model.

\begin{figure}[ht]
        \centering
        \psfig{file=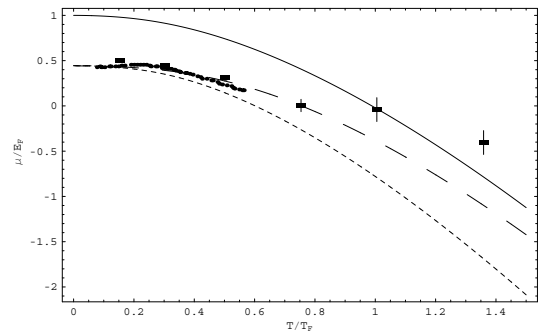,width=7cm,angle=-0}
        \caption{Physical chemical potential versus the recaled temperature.
        The line-styles are similar to Figure.1.
        \small
}\label{fig2}
\end{figure}

With \eq{20} and \eq{11}, we have also shown the chemical potential
versus the rescaled temperature in Fig.\ref{fig2}. The chemical
potential given by the two formalisms decreases with the increase of
temperature. The departure of them is getting bigger with the
increasing temperature.

The results for the energy per particle shown in Fig.\ref{fig1} and
Fig.\ref{fig2} in terms of the two different analytical approaches
are reasonably consistent with the Monte Carlo
calculations\refr{MC1,MC2}, while the chemical potential differs
explicitly from the Monte Carlo result \refr{MC2} for $T/T_{F}>0.8$.

\subsection{Entropy}

\begin{figure}[ht]
        \centering
        \psfig{file=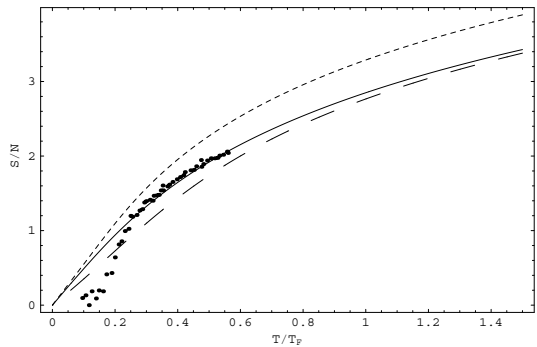,width=7cm,angle=-0}
        \caption{Entropy per particle versus the recaled temperature. The line-styles are similar to
        Figure.1. The Monte Carlo simulation result is extracted from Ref.\cite{MC1}.
        \small
}\label{fig3}
\end{figure}

With \eq{20} and \eq{g}, the entropy per particle curve versus the
    rescaled temperature is presented in Fig.\ref{fig3}.
The quasi-linear approximation predicts that the curve is higher
    than that of the ideal Fermi gas,
    while the generalized exclusion
    statistics model gives lower values compared with that of the ideal Fermi gas.
With the increase of temperature, the entropy per particle given by
the generalized exclusion
    statistics is getting closer to and  almost overlaps with that of the ideal Fermi
gas. In terms of the quasi-linear approximation, the ratio of
    entropy to that of the ideal Fermi gas approaches a constant in the
    Boltzmann regime.

Especially, in the low-temperature strong degenerate regime,
    the slope of the entropy per particle versus the scaled temperature given by
    these two
    approaches is different.
The low-temperature behavior is determined by the effective fermion
    mass according to the Landau theory of strongly
    correlated Fermi-liquid.
In turn, from the entropy curve, one can derive the effective
fermion mass indirectly. The careful study shows that the
quasi-linear approximation indicates $m^*/m\approx  1.11>1$, while
the latter predicts $m^*/m\approx  0.70<1$. Compared with the
latter, the quasi-linear approximation result is more consistent
with the Monte Carlo calculations $m^*/m\sim
1.04-1.09$\refr{Lobo2006,Pilati2008}.

\section{Conclusion}\label{section5}

In terms of the quasi-linear approximation method and generalized
exclusion statistics model, the internal energies, chemical
potentials and entropies of a unitary Fermi gas have been analyzed
in detail. The two different approximations give similar behavior
for the internal energies and chemical potentials of a unitary Fermi
gas.

The entropy is an important characteristic quantity in statistical
mechanics. The entropy by the quasi-linear approximation is higher
than that of the ideal non-interacting fermion gas. In the Boltzmann
regime, the entropy curve given by the generalized exclusion
statistics gets closer towards and almost overlaps with that of the
ideal Fermi gas. The entropy given by the quasi-linear approximation
is getting far away from that of the ideal Fermi gas and the ratio
of entropy to that of the ideal Fermi gas approaches a constant.

According to the quasi-particle viewpoint of the Landau Fermi-Liquid
theory, the slope of entropy per particle determines the effective
fermion mass in the low-temperature strong degenerate region. The
numerical analysis demonstrates  that the generalized exclusion
statistics model gives $m^*/m\approx 0.70<1$. The developed
quasi-linear approximation predicts $m^*/m\approx 1.11>1$, which is
closer to the updating Monte Carlo investigations.

\acknowledgments{ The authors are grateful to
 J.-r Li, X.-w Hou and X.-j Xia for stimulating
discussions. Supported in part by the National Natural Science
Foundation of China under Grant No. 10675052 and 10875050 and MOE of
China under projects No.IRT0624. }

\end{document}